# A Guerilla Approach to Control System Development

J. Dovc, G. Milcinski, M. Plesko, J. Stefan Institute, Ljubljana, Slovenia

## Abstract

We present our experiences in managing the development cycles of the control systems for ANKA and the ALMA Common Software. Our team consists practically only of undergraduate students. Stimulating and rewarding the students with cutting-edge technologies and travel to conferences like this and installation fieldwork are an important positive factor in raising their motivation. However, building any system with a group of inexperienced students is quite a challenging task. Many problems occur with planning deadlines and missing them, organizing and managing development, sources, and documentation and also when dealing with conventional program management rules. To cope with them, we use many tools: CVS for versioning and source archiving, Bugzilla for keeping our bugs in order, a to-do list for managing tasks, an activity log and also many other programs and scripts, some found on the Internet and some made by ourselves. In the end, we had to become organized like a professional company. Documentation and demos can be found on our homepage: http://kgb.ijs.si/KGB. Because of powerful intranet/web front-ends of all those tools, our Internet pages are the central resource for developers, who work mostly off-site.

## 1 INTRODUCTION

The synchrotron light source ANKA has been built at the FZK (Forschungszentrum Karlsruhe), Germany. The control system has been outsourced through a commercial contract to the JSI (J. Stefan Institute) in Ljubljana, Slovenia. There, the KGB group (Kontrol Gruppe für Beschleuniger) has been founded in 1996 by one of the authors of this paper (M.P.) with the very naïve idea that a group of motivated, responsible and skilled students without previous experience in accelerator control systems can build such a system from a ten-pages long wish-list of features and ideas.

## 2 WHY GUERILLA?

It had been clear from the beginning of the ANKA project that the tight project schedule and low budget required an innovative approach both in technology and management. The well-organized groups at the FZK and JSI had experienced engineers who had enough other work to do and were not particularly motivated to take on another project, which did not have clear specifications at that time. We decided that we should bypass rigid organizational structures and use highly motivated people in order to save development time that we were short of.

That called for a guerilla approach (figure 1): Apart from finding people that are waiting to be motivated, sufficiently small tasks had to be defined that could be done in one guerilla action.

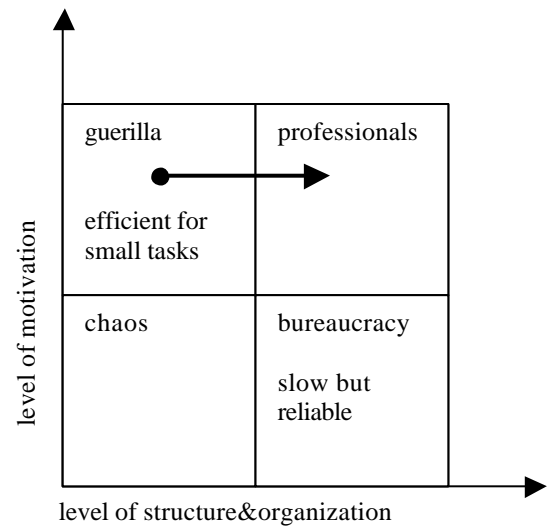

Figure 1: Different organizational forms based on team motivation and organization. We have started at the top left and are moving towards the top right.

### 2.1 The Guerilla Force

The only group of potential team members that might easily be motivated were students we knew from summer jobs. So we started to build our guerilla force, first with one student, then recruiting new members as we met more excellent students of physics, mathematics, electronics and computer sciences with affinity to programming. Whenever the opportunity arose to get a competent person, we took all efforts to ensure that (s)he would come to us. Initially it was not easy to find and to convince people to join us, but it got better as the group had results to show and became sufficiently large that members brought in their friends and classmates.

To make it more compelling for students, we decided to make a control system design based on the Internet and general network technologies, which were becoming very popular at that time like Java, CORBA and the LonWorks fieldbus.

We used any other way of motivating as this was the most important factor in raising the productivity: standard financial compensation, trips to conferences, the chance to work abroad, MP3 music with good speakers and the possibility to use broadband internet access from the institute. Of course, there are no algorithms to deal with people. Each individual is different and the project leader had to adapt his style and rewards to the person and not the other way around, in order to get the most out of that person.

Managing guerillas, who are practically volunteers, is quite challenging in that the leader has no formal leverage over the team. That's why in the end human resource management becomes more important than project management.

## 2.2 The Guerilla Tasks

As the students mostly could not work on a regular basis, sufficiently small tasks had to be made that had to have well-defined and long-term stable interfaces, both in hardware and software. The design was crucial, because a good design allowed to dissect the project in small manageable tasks. It was also important that the assembly of the pieces would not pose extra work. Clear interfaces helped us here.

This concept actually fit well together with the architecture of the control system of ANKA [1]. Because ANKA was a low budget project, it was necessary to offer not just a control system with a low purchase price, but also with:
- low maintenance cost
- low upgrade cost
- low failure rate

i.e., a system with low total cost of ownership (TCO).

In retrospect we can say that there were only two main reasons why we have successfully finished the project. One was horizontal division of tasks and responsibilities: each team member got a specific module to master and to implement, instead of being responsible for the control of a particular kind of equipment, which would correspond to vertical division. Many modules were generic and thus not related to specific equipment, or even to the control system and were used to hide complex details from casual programmers, e.g. the Abeans [2,3] Java-CORBA wrapper or the wizard that generates server code. Actually, most of the time the students didn't even know or care about the details of the controlled devices, whose functionality had been described with the CORBA interface definition language (IDL) according to the wide-interface approach: Each device is represented by a specific CORBA object which has its properties and methods reflecting the state and functionality of the device, (see [3] for more discussion on wide versus narrow interfaces). Since everybody had the guidelines set by all those wide interfaces, there was little that people needed to know to use each other's data as the most information comes from the data themselves. Using this approach, every property has to be well defined in advance and this required a lot of time during the design phase. But this time pays off when programming applications, writing documentation and maintaining the control system. Other benefits of this approach are described in [1].

The other reason for success was pure luck that we got some ingenious students right at the beginning of the project. They programmed the core modular components or developed and programmed the modular I/O boards, respectively, and helped to attract other good students.

## 2.3 Getting the Guerilla Organized

The whole approach probably would not have worked, if we had worked inside of ANKA, because the temptation for continuous improvements, postponing deadlines, not to mention ignoring writing documentation, would be too high. Being responsible to a customer, however, forced us to behave more professionally. We started to implemented several software engineering tools and made our team members use them consistently, which is described in detail in another paper [4]. We have done some improvements, e.g. CVS is now used consistently everywhere, we use documentation templates and UML for designs, but most of the tools are still heavily used.

Apart from inside pressure, outside factors affected our decisions to strengthen our organizational structure. In 1999-2000 we developed a core control system for a commercial company and now, started in 2000, a similar cooperation takes place with ESO (European Southern Observatory) for the ACS (ALMA Common Software).

There were subjects that we never even dared to approach, e.g. C/C++ programming rules. The C++ server was developed by one person, so he had his own rules but at least nobody had others – until we started the project for the commercial company with two new programmers, who couldn't even agree on variable naming, not to mention that the company had a third view. The situation with Java was better as general Java conventions are quite strong. Also, those who used Java learned it after joining the team from the same books and Suns examples and tutorials.

## 2.4 Project Management

Planning was done by the project leader. He also designed the whole control system and distributed the tasks. He used mainly to-do lists by using the features of MS Word in outline view, which we found very useful. The project manager was a charismatic student, whose task was to check the status of the work once a week. Although the productivity of developers increased before the deadline – the effect of last-minute panic, the project manager had to keep reminding them that the deadline was approaching. The most effective solution was to call them one by one, as it was nearly impossible to schedule a meeting with all students attending.

The schedule had to be adjusted to the dynamics of student life (exams, lectures, girl/boyfriends, etc). Most of the work was done during weekends, public holidays, and semester breaks. One of the consequences was that students mostly programmed at home. So we really had to have good means of communication. We used mobile phones (everybody had one), mailing lists for each subproject and weekly meetings.

Initially the hardest work for the project manager was to collect all the corresponding documentation. But by now everybody has learned that documentation is important and some write docs already in advance as specifications. We have close to 1000 pages of manuals for ANKA.

Testing on the other hand was quite successful from the onset. Because the majority of the software we had to test were GUIs, we could not easily use any automatic procedures to test our software. To assure the quality of the products, developers cross-tested the products of each other. That stimulated competition, which is another motivating factor for testing and finding bugs, which we manage using a Web bug reporting and bug tracking tool called Bugzilla.

The main resource of information was the KGB homepage both for the group and for our customers. The publicly available pages contain the complete documentation (manuals, specifications, design documents, white papers, etc.), FAQ, conference articles, presentations, and downloadable demos. The internal pages contain the address book of all the members, archives of mailing lists, and to-do lists. Other Web-based tools that we use heavily are:
- CVS for versioning of the programmers' code and all documentation (also binary formats such as MS Word) and as a repository from which backups are made
- Activity log: a set of simple Perl scripts, where the developers reported their activities
- Bugzilla (http://www.mozilla.org/bugs/)

Although used very efficiently, there is a negative aspect of having all tools on the Web: people who work at home are not online all the time. We are therefore looking for a tool that would cache all entries and synchronize then automatically when going online.

## 3 GUERILLAS TURN PROFESSIONALS

Even after having gotten used to all the project management and software engineering, the team members still are highly motivated. We are using formal document templates, design with UML, talk patterns, apply modular testing and have fun! We have even started JavaAcademy – a training course for newcomers to screen for the best candidates. It appears that we are reaching towards the upper right region in figure 1. Our guerilla approach pays dividends, because it is much easier to get a motivated but chaotic group organized then to motivate an organized team of dull individuals.

So we have a veteran team with an average age of 22. The oldest members have already graduated and left and we would lose all our investment if we let this trend to continue. Our institute cannot hire the whole team; therefore we have created a spin-off company for developing and installing control systems for accelerators and other large experimental facilities (www.cosylab.com). True to the research community that we grew in, the vision of the company is to make a living with our work instead of selling software licenses. And true to our philosophy of high motivation, all initial employees are co-owners.

Now the real management only starts. Will it turn from guerilla to professional? We will know when the company appears at a future ICALEPCS conference as sponsor.

## 4 ACKNOWLEDMENTS

We thank the Forschungszentrum Karlsruhe for placing the initial contract to us, giving us thus the chance to create this excellent team. We also thank the ESO, in particular Gianluca Chiozzi, for teaching us programming and project discipline. We found that it isn't so difficult after all, although Gianluca might argue that we are still far from it.